\documentclass[%
 aip,
 amsmath,amssymb,
 reprint,%
]{revtex4-1}

\usepackage{graphicx}
\usepackage{dcolumn}
\usepackage{bm}
\usepackage{hyperref}
\usepackage[dvipsnames]{xcolor}
\usepackage[utf8]{inputenc}
\usepackage[T1]{fontenc}
\usepackage{mathptmx,mathtools}

\begin{document}

\preprint{AIP/123-QED}

\title[]{Parametric excitation induced extreme events in MEMS and Li\'enard oscillator}

\author{R. Suresh}
\email{sureshphy.logo@gmail.com.}
\author{V. K. Chandrasekar}%
 \email{chandru25nld@gmail.com.}
\affiliation{ 
Centre for Nonlinear Science and Engineering, School of Electrical and Electronics Engineering, SASTRA Deemed University, Thanjavur  613 401, India
}%
\date{\today}

\begin{abstract}
	The two paradigmatic nonlinear oscillatory models with parametric excitation are studied. The authors provide theoretical evidence for the appearance of extreme events (EEs) in those systems. First, the authors consider a well known Li\'enard type oscillator that shows the emergence of EEs via two bifurcation routes: Intermittency and period-doubling routes for two different critical values of the excitation frequency. The authors also calculate the return time of two successive EEs, defined as inter-event intervals, that follow Poisson-like distribution, confirm the rarity of the events. Further, the total energy of the Li\'enard oscillator is estimated to explain the mechanism for the development of EEs. Next, the authors confirmed the emergence of EEs in a parametrically excited microelectromechanical system. In this model, EEs occur due to the appearance of stick-slip bifurcation near the discontinuous boundary of the system. Since the parametric excitation is encountered in several real-world engineering models, like macro and micromechanical oscillators, the implications of the results presented in this paper are perhaps beneficial to understand the development of EEs in such oscillatory systems.
\end{abstract}

\maketitle

\begin{quotation}
	The study of bursting oscillations (large-amplitude oscillations alternated with small-amplitude oscillations) is still an active topic of research due to its ubiquitous nature. Bursting oscillations are encountered and reported in experiments as well as in models of dynamical systems ranging from physics to biology. In contrast with the conventional bursting patterns that occurred in dynamical systems, the development of certain types of oscillatory states, in which the large-amplitude oscillations occasionally appeared in time with an entirely unpredictable nature. These rare and recurrent large-amplitude oscillations are distinguished as extreme oscillations or extreme events (EEs) in the literature. This topic receives significant attention in recent years among the scientific community, and the development of EEs have already been studied and reported in various dynamical systems. In particular, very recently, the emergence of EEs in nonlinear dynamical models influenced by external periodic forcing is reported. Although the emergence of EEs and their mechanism are studied in detail in nonlinear systems with external forcing, very few studies have been reported the appearance of EEs in systems with parametric excitation, in particular, in laser models. However, the influence of parametric excitation to induce EEs in other classes of dynamical systems also requires immediate attention. Therefore, in this paper, the authors provide the theoretical evidence for the appearance of EEs in two paradigmatic nonlinear oscillatory models with parametric excitation. First, the authors consider a well known Li\'enard type oscillator that shows the emergence of EEs via two different bifurcation routes: Intermittency and period-doubling routes for two different critical values of the excitation frequency. Further, the rarity of the EEs is confirmed by calculating the return time of the two successive EEs defined as inter-event intervals that follow the Poisson-like distribution. The total energy of the system is analytically estimated to explain the emerging mechanism of EEs in this oscillator. Next, the authors demonstrated the emergence of EEs in a parametrically excited microelectromechanical system, in which EEs occurred due to the presence of stick-slip bifurcation near the discontinuous boundary of the system.
\end{quotation}

\section{\label{sec1}Introduction}
In dynamical systems, the complex oscillatory patterns with different amplitudes are interspersed. In particular, the coexistence of small and large-amplitude oscillations known as bursting has been encountered and reported in a variety of fields from physics to biology \cite{deschenes1982,medvedev2006,channell2007,deshazer2003}. In contrast with the conventional bursting patterns, there exist intermittent large-amplitude oscillations that occasionally appeared in time are known as extreme oscillations or extreme events (EEs). This type of events occurred in many natural systems and engineering models including oceanography \cite{dysthe2008, donelan2017}, ecosystems \cite{bialonski2015, bialonski2016}, geophysics \cite{sornette2003, ghil2011}, transportation networks \cite{zhao2005, echenique2005, chen2015}, power supply networks \cite{dobson2007}, mechanical oscillators \cite{kingston2017, ksuresh2018,farazmand2019}, neural networks \cite{lehnertz2008, lehnertz2006}, plasma \cite{bailung2011}, optical fiber and lasers \cite{mussot2009, montina2009}, etc and receives notable attention among the scientific community during the past decade. The experimental demonstration of EEs has also been evidenced in many scientific laboratory experiments \cite{kingston2017, solli2007, randoux2012}. Though there is no exact mathematical definition for EEs, according to statistical perspective, it was generally admitted fact that EEs show a long-tailed probability distribution and the peaks which are greater than the threshold height are characterized as EEs. The threshold height is equal to the time-averaged mean value of all the peaks in a measured time series plus 4--8 times the standard deviation derived for the long run. Extreme events with similar statistical properties have already been studied and evidenced in nonlinear dynamical systems modelled by ordinary as well as partial differential equations \cite{ansmann2013, karnatak2014, saha2018, kingston2017, ksuresh2018, kingston2017a, galuzio2014, rothkegel2014, clerc2016, chang2015, suresh2018}. Specifically, in refs \cite{kingston2017,suresh2018,ksuresh2018}, the authors have reported the influence of external periodic force to induce EEs in a Li\'enard type oscillatory model and microelectromechanical system (MEMS) model.

In general, an external periodic force might affect the oscillatory system, either additively or multiplicatively. In the former case, the entire system is driven by an external force, and in the latter situation, the periodic force influences on any one of the system parameters, yielding quite different types of solutions. For example, the slow reduction of the catalytic activity in chemical reactions due to chemical erosion decreases the reaction performance \cite{elizalde2014}. Periodic modulation in one of the system parameter in the microelectromechanical device induces the parametric resonance, which is used as vibration energy harvesters \cite{jia2014, jia2013, jia2016}. Electrostatically driven microelectromechanical systems are used to design highly effective bandpass filters \cite{rhoads2005}. In all these examples, there are certain control parameters of the system vary periodically as a function of time or manually altered between a specific range. 
Further, slowly varying parameters can lead to unusual and counter-intuitive effects such as stabilizing the unstable fixed points in the inverted pendulum \cite{,schiele1997,champneys2012}, periodic delay bifurcation in nonlinear dynamical systems \cite{lakrad2004}, etc. Therefore, the study of parametric excitation or slowly varying control parameters in nonlinear dynamical systems have been and continued to be an active topic of research in many fields (See the refs \cite{yue2014, han2015, han2018, han2018a, han2018b} and the references therein). 

Even though the study of EEs and their emerging mechanism in the systems with external forcing are studied and acknowledged by the scientific community, very few studies have been reported the emergence of EEs in parametrically excited nonlinear dynamical systems, especially in laser models\cite{ksuresh2018,metayer2014,bonatto2017,gomel2019}. However, the influence of parametric excitation to induce EEs in other classes of dynamical systems, in particular, macro and micromechanical oscillator models also require immediate attention. Therefore, in the present study, the authors aim to investigate the dynamical changes that occurred in the nonlinear systems in response to the parametric excitation. In particular, they report the evidence of the occurrence of EEs in two paradigmatic examples of nonlinear oscillators, a Li\'enard type model and in a cantilever-based MEMS model with parametric excitation. When one first considers the Li\'enard oscillator, for the suitable parameter values, the large-amplitude oscillations appeared via two bifurcation routes, namely intermittency and period-doubling routes at two critical values of the excitation frequency. The threshold height is estimated to classify the EEs from the frequent large-amplitude oscillations. In addition to that, the return time of the two consecutive EEs is calculated, known as inter-event intervals, which follows a Poisson-like distribution confirming the rare occurrence of the events. Further, the authors explained the mechanism responsible for the development of EEs in the Li\'enard oscillator using the total energy of the system. The occurrence of parametric excitation induced EEs is robust against system dynamics. To verify this, next, the authors consider a MEMS model with discontinuous boundaries \cite{ksuresh2018} to demonstrate the parametric excitation induced EEs. The models with discontinuous boundaries are frequently encountered in mechanical oscillators like cantilever-based microelectromechanical oscillators \cite{liu2004, evans2014}, mass-spring-damper oscillator \cite{geffert2017}, systems with friction \cite{li2016}, etc. Furthermore, the authors show that the existence of stick-slip bifurcation near the discontinuous boundary region is the underlying mechanism for the appearance of EEs in the MEMS model.

One can also note here that, although when one implements the parametric excitation in the internal frequency of the Li\'enard and MEMS models, the underlying mechanism for the emergence of EEs is different for both the systems. Furthermore, usually, the parametric resonance occurs when the external excitation frequency equals to any integer multiples of the natural frequency of the oscillator. In both the systems, the emergence of EEs appeared for a specific region of the excitation frequency. However, the authors could not find any relation between them in the present study. 

The remainder of this paper has been organized as follows: In Sec.~\ref{sec2}, the authors introduce the Li\'enard oscillator with parametric excitation and demonstrate the emergence and mechanism of EEs. Section~\ref{sec3} is devoted to the study of EEs in a parametrically excited MEMS model. Finally, in Sec.~\ref{sec4}, the authors summarize their results with conclusions. 
\section{\label{sec2}Extreme events in the Li\'enard oscillator}
\subsection{\label{sec2a}Dynamical model}
To start with, first, the authors consider a specific class of Li\'enard-type nonlinear oscillator model with parametric excitation; that is,
\begin{align}
\ddot{x}+\alpha x\dot{x}-\gamma [1+F \cos(\omega t)]x +\beta x^{3} = 0,
\label{eq1} 
\end{align}
where $\alpha$ is the magnitude of the position-dependent damping or nonlinear damping, $\gamma$ is the internal frequency parameter of the oscillator, and $\beta$ represents the strength of cubic nonlinearity. The parametric excitation amplitude and frequency are represented by $F$ and $\omega$, respectively. In this model, the parametric excitation is introduced in the internal frequency of the system. Thus, the intrinsic frequency parameter $\gamma$ is periodically time-varying as a function of $F$ and $\omega$.

Equation (\ref{eq1}) is equivalent to a well-known Mathieu's equation\cite{poulin2008} when the damping is linear and $\beta=0$. Mathieu's equation exhibits unstable and stable behavior under parametric excitation, which manifests into EEs in the stochastic generalization \cite{mohamad2016}. Further, the mechanism causing EEs is also evident in the stochastic case\cite{mohamad2015}. Nevertheless, in the present study, the authors have considered a deterministic nonlinear equation (\ref{eq1}) with parametric excitation and demonstrated the emergence of EEs as a function of excitation parameters. When $F=0$ in Eq.~(\ref{eq1}), the system can be viewed as a cubic anharmonic oscillator with nonstandard Hamiltonian nature\cite{chandru2005}, or as a conservative nonlinear oscillator perturbed by a nonlinear damping term ($\alpha x\dot{x}$). This type of oscillatory model with nonlinear damping arises in a broad class of physical, mechanical, chemical systems and engineering models with appropriate transformations. For instance, various micro and nanoelectromechanical systems having nonlinear damping coefficient in the form of restoring force or stiffness coefficient\cite{eichler2011}, likely the nonlinear elements in the electronic circuits. Besides, the autonomous Li\'enard oscillator shows the bistable nature of the coexistence of dissipative and conservative dynamics depending on the initial conditions \cite{chandru2005b,mishra2015}. Hence, the trajectories either dissipate and approach to the stable fixed point or exhibit self-sustained periodic oscillations subject to the initial conditions. Thus, the nonlinear damping term in the Li\'enard oscillator act as damping, and also pumping term (which gives rise to self-sustained oscillations) depending on the amplitude of the oscillations.

Further, it is worth to emphasize that the autonomous Li\'enard oscillator is an example of the reversible system under the transformation $S: x\rightarrow-x$, $t\rightarrow -t+T$, where $T (\text{period})=\frac{2\pi}{\omega}$. Due to this reversible property, Li\'enard oscillator plays an important role in Hamiltonian and non-Hamiltonian dynamics. One can also note here that the Li\'enard oscillator has isochronous oscillation property and exhibiting parity-time ($\mathcal{PT}$)-symmetric nature \cite{karthiga2016}. In the next section, the dynamics of Eq.~(\ref{eq1}) will be studied in detail in terms of the excitation parameters. The development and generation mechanism of EEs will also be explained. 
%
\begin{figure}[tbp]
	\centering
	\includegraphics[width=1.0\columnwidth]{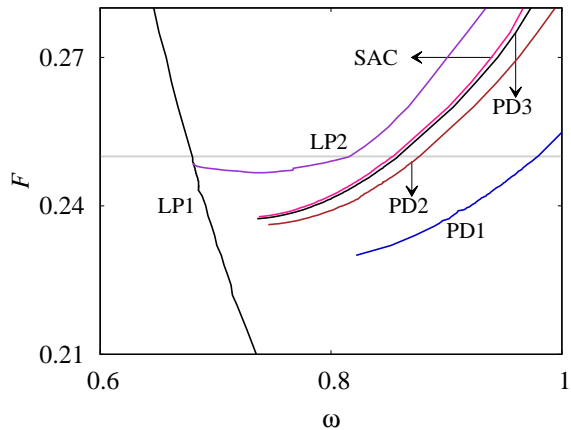}
	\caption{The two-parameter bifurcation diagram of the Li\'enard oscillator (\ref{eq1}) in the ($F, \omega$) plane depicts the emergence of different bifurcations. The system parameters are fixed as follows: $\alpha = 0.44$, $\gamma = 0.5$, and $\beta = 0.5$. For other details, see the text.} 
	\label{fig1}
\end{figure}
\begin{figure}[tbp]
	\centering
	\includegraphics[width=1.0\columnwidth]{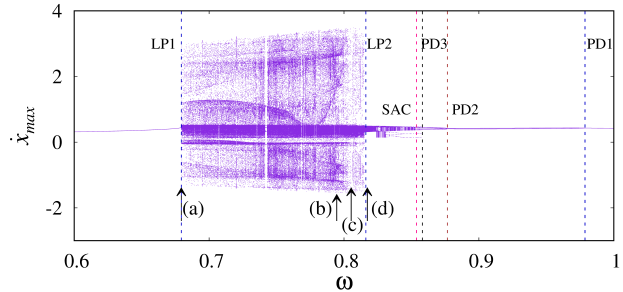}
	\caption{The maxima of the dynamical variable $\dot{x}$ of the oscillator (\ref{eq1}) are plotted against the parametric excitation frequency in the range of $\omega\in(0.6, 1.0)$ with the fixed value of $F=0.25$ (where the horizontal line is marked in Fig.~\ref{fig1}). The other system parameters are fixed as in Fig.\ref{fig1}. The vertical dashed lines indicate the $\omega$ values where different bifurcations occur. The arrow marks and the corresponding labels from (a) to (d) represents the $\omega$ values where the time evolution plots are plotted which are depicted in Fig.~\ref{fig3}.} 
	\label{fig2}
\end{figure}
\subsection{\label{sec2b}Results and Discussion}
The dynamics of an autonomous Li\'enard oscillator and its dual characteristic nature of dissipative and conservative dynamics are well studied and reported. Further, the effect of external periodic forcing in the Li\'enard oscillator is examined, and the appearance of periodic mixed-mode oscillations and EEs in the model with external periodic forcing has also been recently evidenced with experimental confirmations \cite{kingston2017,kingston2017a, suresh2018}. In the present paper, the authors have considered a Li\'enard oscillator model. Still, instead of the external force driving the system, it acts only on the internal frequency ($\gamma$), resulting in the periodic change in $\gamma$.

Equation (\ref{eq1}) is numerically integrated using the fourth-order Runge-Kutta method with the time step of $0.01$, and for the present numerical study, the system parameters are fixed as $\alpha=0.44$, $\gamma=0.5$, and $\beta=0.5$. The initial conditions are chosen from the dissipative region. Therefore, the trajectories of the autonomous oscillator damp and approach to the stable fixed-point, or the forced oscillator exhibit periodic or non-periodic solutions based on the values of the excitation parameters. The system dynamics is examined by varying the excitation amplitude in the range of $F\in[0.21,0.28]$ and frequency $\omega\in[0.6,1.0]$. The two-parameter bifurcation is depicted in Fig.~\ref{fig1} to show the occurrence of various bifurcations. Further, in Fig.~\ref{fig2}, the qualitative changes that occurred in the Li\'enard oscillator at different bifurcation points are portrayed as a one-parameter bifurcation diagram for $\omega\in[0.6,1.0]$ by fixing $F=0.25$ (at which the horizontal line is marked in Fig.~\ref{fig1}). The curves marked as PD1, PD2 and PD3 in Fig.~\ref{fig1} represents the parameter values at which the period-doubling cascades occur. These states are manifested in Fig.~\ref{fig2} by the vertical dotted lines for $\omega=0.9786$ (PD1), $0.8766$ (PD2) and $0.8580$ (PD3), respectively. The period-doubling cascades then lead the system to exhibit chaotic dynamics when one varies the excitation parameters. The emerging small-amplitude chaos is indicated as SAC in Fig.~\ref{fig1}. The onset of small-amplitude chaotic attractor via period-doubling bifurcation is marked in Fig.~\ref{fig2} as SAC for $\omega=0.8538$. When one further decreases the excitation frequency, the system size is suddenly increased due to the interior crisis, in which the small-size chaotic attractor suddenly bifurcates into a large-size chaotic attractor. The curve LP2 in Fig.~\ref{fig1} denotes the set of parameters at which an interior crisis occurred. This dynamical transition can be visualized from Fig.~\ref{fig2} for $\omega=0.8153$ at which the small-sized attractor explodes into a large-sized chaotic attractor consist of occasional large-amplitude oscillations alternates with the small-amplitude chaotic oscillations. One can note here that crisis is a common manifestation of chaotic dynamics and EEs which have been observed in many experimental and theoretical studies \cite{kingston2017,grebogi1982,grebogi1987,eschenazi1989,bonatto2017}.
Further, the amplitude of the large-sized attractor is slowly decreasing in size with a reduced value of $\omega$ and when $\omega\approx 0.67934$, the large chaotic attractor suddenly transformed into a periodic attractor via intermittency bifurcation route (marked as LP1 in Figs.~\ref{fig1} and \ref{fig2}). Therefore, from Fig.~\ref{fig2}, one can realize that the emergence of large-amplitude oscillations occurred via two bifurcations routes, one is from the period-doubling route when decreasing the excitation frequency from higher values and the second route emerged via intermittency when one increases $\omega$ from lower to higher values. The authors emphasize here that the intermittency route to the onset of EEs have already been reported in several other dynamical models such as neuron models, in semiconductor and optically injected laser systems\cite{mishra2018,reinoso2013,munt2013}.

In a nutshell, it is evident from Fig.~\ref{fig2} that the chaotic dynamics have emerged through two distinct routes at two different critical values of the excitation frequency. Also, it is clear from Fig.~\ref{fig2} that for $\omega\in(0.67934, 0.8153]$, the Li\'enard oscillator exhibit large-sized chaotic attractor in which the large-amplitude chaotic oscillations alternates with the small-amplitude chaotic oscillations in the time domain. Specifically, for a certain range of $\omega$, the large-amplitude oscillations occurred occasionally and at random time intervals, which are then characterized as EEs. On the other hand, for other values of $\omega$, the system exhibits frequent large-amplitude chaotic oscillations. Therefore, to distinguish EEs, the authors have used a threshold height, first proposed by Massel\cite{massel1996}to define extreme rouge waves that occurred in oceans, which has then been widely used to characterize EEs in the literature of extreme value theory in recent times\cite{dysthe2008,karnatak2014,ansmann2013,kingston2017,suresh2018,kharif2009,bonatto2011,ray2020}. The threshold height can be defined from the dynamical aspect as a deviation of several standard deviations away from the average value of the system observable, given as,
\begin{align}
H_T=\langle P\rangle+n\sigma,
\label{eq1a}
\end{align} 
where $\langle P\rangle$ is the time-averaged positive peak value of the $\dot{x}$ component of the Li\'enard oscillator, $\sigma$ is the standard deviation of the $\dot{x}$ variable, and $n$ is an integer, which is system-dependent. For the Li\'enard oscillator, the value of $n$ is chosen as $8$ \cite{kingston2017,chowdhury2019}. In order to calculate the threshold height $H_T$, Eq.~(\ref{eq1}) is numerically integrated for a long run with the iterations of $2\times10^{9}$ time units of the system variable $\dot{x}$, after leaving sufficient transients. During the emergence of frequent large-amplitude oscillations, the average value of the large peaks is very high, and so $H_T$ becomes larger than the largest peaks. On the other hand, the occasional occurrence of large peaks whose amplitude is larger than $H_T$ are identified as EEs.
\begin{figure}[tbp]
\centering
\includegraphics[width=1.0\columnwidth]{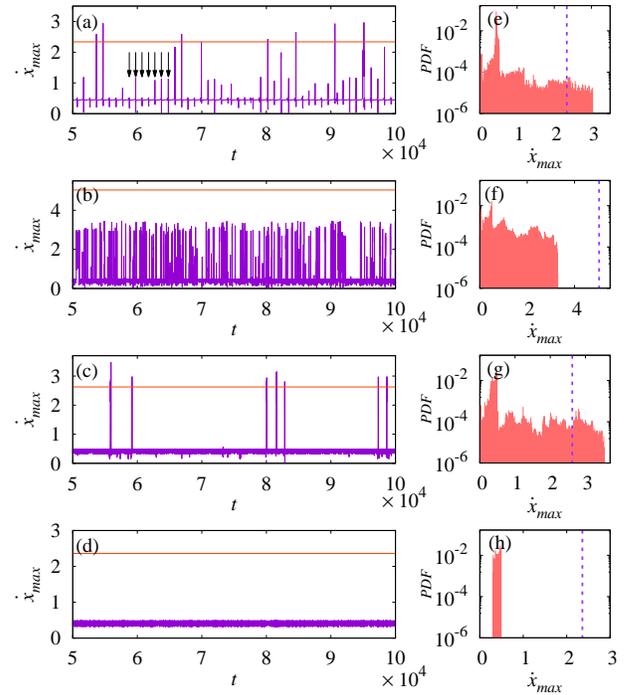}
\caption{(a) - (d) Temporal evolution of the maxima of $\dot{x}$ variable ($\dot{x}_{max}$), and (e) - (h) their respective probability distribution functions (PDF) of the oscillator (\ref{eq1}) depict various dynamical states observed in the Li\'enard oscillator for different values of $\omega$. (a) and (e) shows EEs emerged from the intermittency route for $\omega=0.6794$, (b) and (f) indicate the emergence of very frequent large-amplitude oscillations for $\omega = 0.7946$, (c) and (g) shows EEs developed via period doubling route for $\omega = 0.8053$ and (d) and (h) represents the bounded small-amplitude chaotic oscillations for $\omega = 0.8154$. The horizontal line in (a) -- (d) and dashed vertical line in (e) -- (h) depict the threshold height $H_T$ and the long-tail behavior of the PDF in (e) and (g) corroborate the occurrence of EEs.} 
\label{fig3}
\end{figure}

The temporal evolution of the $\dot{x}$ variable of the Li\'enrd system is plotted in Figs.~\ref{fig3} (a) - (d) for different values of excitation frequency $\omega$. Here, the authors have plotted the maxima of $\dot{x}$ variable ($\dot{x}_{max}>0$) for better clarity. The red (dark gray) horizontal line in these figures represent the value of $H_T$. For an illustration, if one notices the bifurcation diagram in Fig.~\ref{fig2} with increasing $\omega$, the periodic orbit suddenly bifurcates into a large-sized chaotic attractor at $\omega\approx 0.67934$ via intermittency route. At this value of $\omega$, the system displays a combination of small-amplitude oscillations along with occasional intermittent large-amplitude chaotic bursts. The temporal evolution of the system for $\omega = 0.6794$ is shown in Fig.~\ref{fig3}(a), where there exist small-amplitude peaks along with intermittent large-amplitude chaotic bursts, in which few of the large peaks are qualified as EEs since they are larger than the threshold $H_T$. If one closely look into Fig.~\ref{fig3}(a), the bursting or spikes (arrows mark some of them) have occurred almost at periodic intervals. Nevertheless, the time difference between the two successive EEs occurred at random intervals of time. Further, the authors also estimated the probability distribution function (PDF) of the time series (Fig.~\ref{fig3}(a)), which shows the long-tail behavior (trademark property of EEs). The PDF is plotted in Fig.~\ref{fig3}(e) and the vertical dashed line indicates the value of $H_T$.
\begin{figure}[tbp]
	\centering
	\includegraphics[width=1.0\columnwidth]{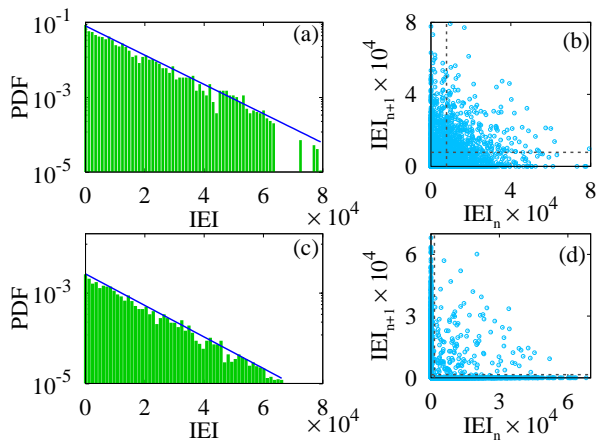}
	\caption{(a) and (c) Inter-event interval (IEI) histogram of the Li\'enard oscillator shows the Poisson-like distribution confirming the rare nature of the EEs for $\omega=0.6794$ and $\omega=0.8153$, respectively. (b) and (d) shows the corresponding joint inter-event intervals. Each open circle demotes a spike. The mean value of JIEI is shown as a dashed line.} 
	\label{fig4}
\end{figure}

Further, when one starts increasing the excitation frequency ($\omega$) for higher values, the occurrence of large-amplitude oscillations is also increasing, and the switching from a small-amplitude oscillation to large-amplitude oscillation is very frequent. The temporal dynamics of such a dynamical state is illustrated in Fig.~\ref{fig3}(b) for $\omega=0.7946$, where the large-amplitude bursts transpire very frequently. Therefore, the threshold height $H_T$ has a large value than the largest peaks, and no events are qualified as EEs, which is also evident from Fig.~\ref{fig3}(b). The corresponding PDF of Fig.~\ref{fig3}(b) is plotted in Fig.~\ref{fig3}(f), which almost has an equal number of large events compare to the small events. Figure ~\ref{fig3}(c) is plotted for $\omega = 0.8053$ exemplify the occasional appearance of large-amplitude chaotic bursts in which few of the peaks are higher than $H_T$ denotes the EEs dynamics. Moreover, one can also observe that unlike in Fig.~\ref{fig3}(a), here (in Fig.~\ref{fig3}(c)) the time difference between successive bursts (events) appeared completely at random intervals of time. The long-tail behavior of PDF (Fig.~\ref{fig3}(g)) once again confirming the occurrence of EEs. Beyond $\omega=0.8153$, the system displays small-amplitude chaos and the corresponding time evolution and PDF are plotted in Fig.~\ref{fig3}(d) and Fig.~\ref{fig3}(h), respectively for $\omega=0.8154$.

The rarity and random occurrence of EEs are further quantified by calculating the statistical properties of events such as the distribution of inter-event interval (IEI) and joint inter-event interval (JIEI) histograms, which are adopted from the literature of biological spiking neurons\cite{chen2009,fitzurka1999}. The IEI sequences (IEI$_n$) is defined as the history of time intervals between consecutive events in the event train. The authors considered those bursts as events that are higher than the threshold height $H_T$. Let $t_n$ be the occurrence time of the $n$th event in a set of $N$ events, then the IEI$_n$ is a variable:
\begin{equation}
IEI_n = t_{n+1}-t_n, \quad n = 1,2,\cdots,(N-1).
\end{equation}
The JIEI histogram can be defined as the serial correlation of IEI following an event (IEI$_{n+1}$) as a function of the proceeding one (IEI$_n$). To estimate the IEI and JIEI, the authors have numerically generated large data ($2\times10^{9}$ time units) by leaving sufficiently large transient. The PDFs of the IEI are plotted in Figs.~\ref{fig4}(a) and \ref{fig4}(c) using a semi-log scale, corresponding to the time evolution plots presented in Figs.~\ref{fig3}(a) and \ref{fig3}(c), for $\omega=0.6794$ and $\omega=0.8053$, respectively, which follows Poisson-like distribution corroborating the rarity of the events. One can note here that the probability of a short IEI occurrence is much greater than a longer IEI occurrence. The PDFs are then fitted by $P(x)=\lambda e^{-\lambda x}$, where $x$ is the inter-event interval and $\lambda>0$ is a scaling parameter. The respective Poisson distributions are fitted with their corresponding IEI by blue (dark gray) lines depicted in Figs.~\ref{fig4}(a) and \ref{fig4}(c) with the value of scaling parameter $\lambda=0.000925$ and $\lambda=0.0000837$, respectively. This Poisson-like distribution is further confirmed by plotting the JIEI histograms in Figs.~\ref{fig4}(b) and \ref{fig4}(d) for the same values of $\omega$. The dashed lines in  Figs.~\ref{fig4}(b) and \ref{fig4}(d) indicates the mean $\displaystyle\sum_{n=1}^{N-1}\text{IEI}_{n}/(N-1)$ value of JIEI. The JIEI analysis reveals that most of the event intervals are found to fall close to the origin of $x$- and $y$-axis, indicating that most of the time there is little change in consecutive event intervals. When one moves away from the origin, the points in the JIEI histogram follows a negative exponential distribution.
\begin{figure*}[tbp]
	\centering
	\includegraphics[width=2.0\columnwidth]{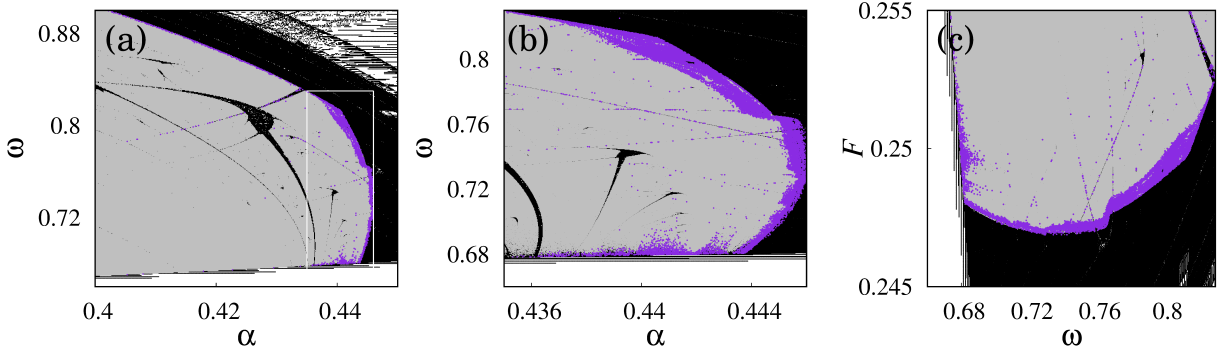}
	\caption{(a) The two-parameter diagram in the ($\alpha, \omega$) plane of the parametrically excited Li\'enard oscillator (\ref{eq1}) shows the occurrence of different types of oscillatory states for $F=0.25$. Other parameters are fixed as given in Fig.~\ref{fig1}. The light-gray represents the region where very frequent oscillations occurred, purple (dark-gray) indicates the EE region, the black domain illustrates the region of small-amplitude chaotic oscillations and the white area depict the periodic oscillations. (b) Magnified area of Fig.~\ref{fig5}(a) clearly shows the occurrence of EEs (purple points). (c) Two-parameter diagram of the Li\'enard oscillator in ($F, \omega$) plane with $\alpha=0.44$ shows the occurrence of different oscillatory states.} 
	\label{fig5}
\end{figure*}

The global dynamics of the oscillator (\ref{eq1}) as a function of $\alpha$, $F$ and $\omega$ is portrayed as a two-parameter diagram depicted Fig.~\ref{fig5}. In particular, in Fig.~\ref{fig5}(a) the authors have varied the parameters $\alpha\in[0.4,0.45]$ and $\omega\in[0.65,0.9]$ on a fine grid and integrated the system for a long run for $F=0.25$. Different dynamical states are separated as follows: The peaks which are above the sub-threshold value of the system observable $\dot{x}=1$ are considered as large-amplitude oscillations. Further, to distinguish EEs from the frequent large-amplitude oscillations, the threshold height $H_T$ is used. Here, the value of $H_T$ is calculated at each set of parameters. More specifically, for the set of parameters to which the peak values of the oscillator are larger than $H_T$ are marked as EEs region. Smaller than $H_T$ is identified as the region with frequent large-amplitude oscillations. Furthermore, the chaotic region is separated from the periodic state by estimating the largest Lyapunov exponent of the Li\'enard oscillator for each set of parameters. The purple (dark gray) region in Fig.~\ref{fig5}(a) indicates the EEs, light gray represents the region of very frequent large-amplitude oscillations. The black region stands for the small-amplitude chaos, and finally, the white area indicates the periodic oscillations. The enlarged area of Fig.~\ref{fig5}(a) (marked as a rectangle) is depicted in Fig.~\ref{fig5}(b) in which EEs region is manifested. Similarly, the same dynamical states as a function of excitation amplitude ($F$) and frequency ($\omega$) is also identified and portrayed with the same color codes in Fig.~\ref{fig5}(c) by fixing $\alpha=0.44$. From Fig.~\ref{fig5}, one can notice the two different routes (period-doubling and intermittency) to the emergence of EEs for a wide parameter range. If one fixes $\alpha$ in Fig.~\ref{fig5}(a) or $F$ in Fig.~\ref{fig5}(c) for a constant value and vary the excitation frequency $\omega$, the emergence of EEs is perceived via intermittency and period-doubling routes.

In the next section, the authors will provide analytical reasoning and mechanism for the development of EEs in the Li\'enard oscillator.
\subsection{\label{sec2c}Possible Mechanism for EEs}
\begin{figure}[tbp]
	\centering
	\includegraphics[width=1.0\columnwidth]{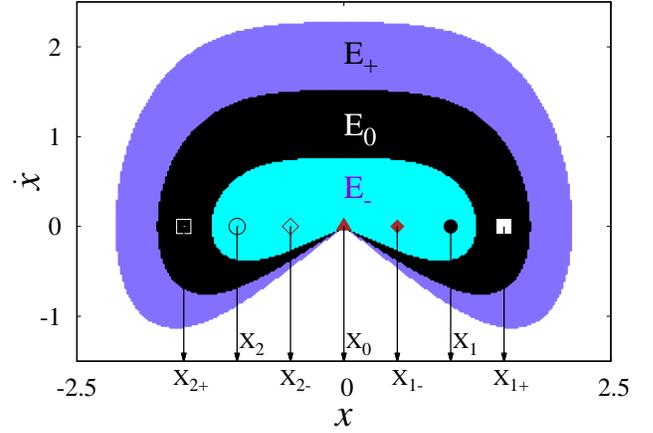}
	\caption{The phase-space diagram of the Li\'enard oscillator (\ref{eq1}) depicts the dissipative regions for three different situations. The black region ($E_0$) represents the dissipative region for $F=0$ in Eq.~(\ref{eq1}). purple (dark gray) area ($E_+$) indicates the dissipative region of the oscillator when the excitation amplitude is maximum ($+F$), and the aqua (light gray) region ($E_-$) represents the dissipative region when the excitation amplitude is minimum ($-F$). The arrows represent the position of the equilibrium points of the Li\'enard oscillator for the above three cases that are depicted in different symbols and labels. For more details, see the text.} 
	\label{fig6}
\end{figure}

For $F=0$ in Eq.~(\ref{eq1}), the Li\'enard oscillator has three equilibrium points $X_{0}=(0,0)$, $X_{1}=\left(+\sqrt{\frac{\gamma}{\beta}},0\right)$ and $X_{2}=\left(-\sqrt{\frac{\gamma}{\beta}},0\right)$. For the chosen parameter values (as in Sec.~\ref{sec2a}), the system has a saddle point at $X_{0}$ = (0, 0), a stable focus at $X_{1}$ = (1, 0) and an unstable focus at $X_{2}$ = (-1, 0). All these three fixed points are illustrated in Fig.~\ref{fig6} as filled triangle ($X_{0}$), filled circle ($X_{1}$), and an open circle ($X_{2}$), respectively. Interestingly, due to the presence of the nonlinear damping term ($\alpha x\dot{x}$) in Eq.~(\ref{eq1}) the stable focus at (1, 0) is linearly stable and nonlinearly unstable. Hence, the trajectories dissipate and approach to the stable focus if one chooses the initial conditions within some region of the phase-space. Otherwise, the system exhibit nonisochronous periodic oscillations. Therefore, based on the choice of initial conditions the system has either dissipative or conservative nature. One can examine the dual nature of the Li\'enard oscillator in terms of the total energy of the system. The total energy of the oscillator (\ref{eq1}) without parametric excitation ($F=0$) can be written as \cite{suresh2018}
\begin{eqnarray}
{E_{0}} = &\frac{1}{2}\left[\dot{x}^{2}+\frac{\alpha \dot{x}}{2}\left(x^{2}-\frac{\gamma}{\beta}\right)+\frac{\beta}{2}\left(x^{2}-\frac{\gamma}{\beta}\right)^{2}\right]\nonumber \\
& \times ~e^{\frac{\alpha}{\Omega}\tan^{-1}\left[\frac{\alpha \dot{x}+2\beta\left(x^{2}-\frac{\gamma}{\beta}\right)}{2\Omega \dot{x}}\right]}-\left(\frac{\gamma^2}{4\beta}\right)e^{\frac{\alpha\pi}{2\Omega}},
\label{eq1b} 
\end{eqnarray}
where $\Omega=\left(\frac{1}{2}\right)\sqrt{8\beta-\alpha^2}$. For $\gamma\geq0$ and $8\beta>\alpha^{2}$, the system displays the dual nature of conservative and dissipative dynamics even if the system admits Hamiltonian. If one substitutes the initial conditions for ($x,\dot{x}$) in Eq.~(\ref{eq1b}), for some initial conditions $E_0$ has negative values and for those initial conditions the system exhibit dissipative dynamics. On the contrary, the system has conservative dynamics, when the total energy of the system $E_0\ge0$. The boundary which separates the dissipative and conservative region is called homoclinic orbit.
The black region in Fig.~\ref{fig6} displays the total energy of the oscillator when $E_0<0$ and so the trajectories originated from this domain have dissipative nature and converge towards the stable focus at (1,0) when time persists.

In the presence of parametric excitation ($F\neq0$), the fixed points of the system start oscillating along the x-axis with respect to the excitation forcing $F\sin(\omega t)$. To be more specific, the stable and unstable equilibrium points move symmetrically back and forth as a function of time (but in opposite directions), and the saddle point $X_{0}$ remains the same. For example, the stable equilibrium point $X_{1}$ oscillates from $X_{1+}=\left(+\sqrt{\frac{(\gamma+F)}{\beta}},0\right)$ to $X_{1-}=\left(+\sqrt{\frac{(\gamma-F)}{\beta}},0\right)$ and the unstable fixed point $X_{2}$ oscillates from $X_{2+}=\left(-\sqrt{\frac{(\gamma+F)}{\beta}},0\right)$ to $X_{2-}=\left(-\sqrt{\frac{(\gamma-F)}{\beta}},0\right)$. These points are marked in Fig.~\ref{fig6} as a filled square ($X_{1+}$), open square ($X_{2+}$), filled diamond ($X_{1-}$) and open diamond ($X_{2-}$) points for $F=0.25$. Due to the movement of the equilibrium points, the dissipative and conservative regions of the oscillator oscillates in the time domain as a function of the excitation amplitude and frequency. Consequently, the dissipative region shrinks and expand in the phase-space like a balloon depending on the value of $F\sin(\omega t)$. To confirm this, the total energy of the oscillator (\ref{eq1}) is again computed for maximum ($+F$) and minimum ($-F$) values of $F\sin(\omega t)$ which is given by
\begin{eqnarray}
{E_{\pm}} = &\frac{1}{2}\left[\dot{x}^{2}+\frac{\alpha \dot{x}}{2}\left(x^{2}-\frac{(\gamma\pm F)}{\beta}\right)+\frac{\beta}{2}\left(x^{2}-\frac{(\gamma\pm F)}{\beta}\right)^{2}\right]\nonumber \\
& \times ~e^{\frac{\alpha}{\Omega}\tan^{-1}\left[\frac{\alpha \dot{x}+2\beta\left(x^{2}-\frac{(\gamma\pm F)}{\beta}\right)}{2\Omega y}\right]}-\left(\frac{(\gamma\pm F)^2}{4\beta}\right)e^{\frac{\alpha\pi}{2\Omega}}.
\label{eq1c} 
\end{eqnarray}
The dissipative dynamics of $E_+$ and $E_-$ regions are depicted in Fig.~\ref{fig6} as purple (dark-gray) and aqua (light-gray), respectively. The system shows chaotic dynamics when it satisfies the condition 
\begin{eqnarray}
F\gtrapprox \frac{\alpha  \gamma ^{3/2} \sqrt{\frac{\gamma }{\beta }} \sinh \left(\frac{\pi  \omega }{2 \sqrt{\gamma }}\right)}{2 \sqrt{2} \omega ^2},
\label{eq1c} 
\end{eqnarray}
which was analytically derived from Melnikov's method \cite{han2012a} by assuming the nonlinear damping ($\alpha$) and the excitation ($F$) terms in Eq.~(\ref{eq1}) as perturbations. For the suitable values of excitation parameters, the oscillator exhibits small-amplitude chaotic oscillations, which is confined in the small area of the phase-space for a long time. During the expansion and contraction of the dissipative region, the system trajectories also crossing the dissipative region near the saddle point, at irregular time intervals, is then strongly repelled towards the unstable direction into the conservative region. Therefore, the trajectories gain energy and make large excursions in the phase-space resulting in large-amplitude oscillations. The trajectories which are making large excursions are then return to the dissipative region after a while and confined in the small-amplitude attractor. The next large excursion is possible only when the above stated condition occurred again.

It is important to emphasize that when one includes an external force as an additive term into the Li\'enard oscillator, then the movement of the equilibrium points in the phase-space is entirely different\cite{suresh2018} than the present case. For instance, if the forcing amplitude ($F$) has a positive (negative) value, then the fixed points $X_2$ ($X_1$) and $X_0$ move towards each other. When $F$ reaches the maximum (minimum) threshold, the $X_2$ ($X_1$) and $X_0$ equilibrium points collide each other and disappeared via saddle-node bifurcation and only $X_1$ ($X_2$) fixed point is feasible. Hence, the three-fixed-point system is transformed into a single-fixed-point system and vice versa as a function of time. Nevertheless, as discussed earlier, when one includes the forcing as a multiplicative term in the Li\'enard oscillator, then $X_1$ and $X_2$ equilibrium points move symmetrically back and forth as a function of time (but in opposite directions) and $X_0$ remains in the same position. Thus, the movement of equilibrium points and the emerging mechanism of EEs in the Li\'enard oscillator with the addictive forcing term is different from the mechanism discussed in the present paper.

Considering the current problem, one can also note that the expansion and contraction of the dissipative region occurred periodically regarding the excitation frequency.  Whereas, the chaotic dynamics of the system depends on both excitation and internal frequency of the oscillator (in this study, the internal and external excitation frequencies are dissimilar). The large excursions only occurred when these two incidents coexist at the same time near the saddle point. If these two events coexisted at infrequent intervals, then the oscillator exhibit EEs. Otherwise, the large-amplitude oscillations occurred more frequently.

\begin{figure}[tbp]
	\centering
	\includegraphics[width=1.0\columnwidth]{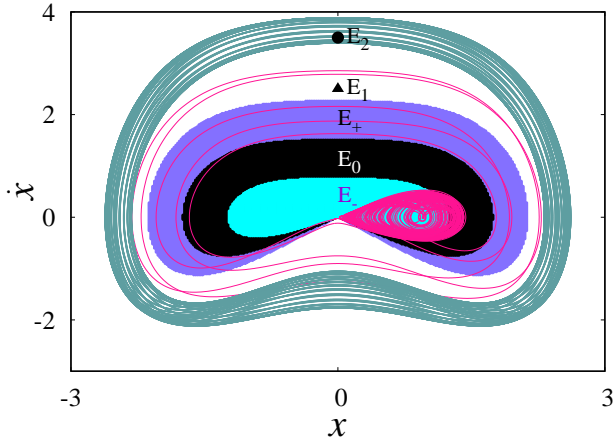}
	\caption{Evolution of the trajectories of the Li\'enard oscillator when one chooses the initial conditions in different energy regions. The chosen initial conditions are represented as a filled triangle and the filled circle. The dissipative regions depicted in Fig.~\ref{fig6} for different values of $F$ are plotted here for comparison. The chaotic dynamics of the oscillator is plotted as pink (light gray) lines when it is started from very near to the region $E_{+}$. The oscillator exhibits quasiperiodic dynamics when one chooses the initial conditions in the $E_{2}$ region (away from $E_{1}$ region).} 
	\label{fig7}
\end{figure}

As mentioned earlier, the oscillator has bistable nature based on the choice of initial conditions. If a trajectory is originated from the initial condition that is very near to the dissipative region with energy $E_1$ (which is near, but greater than $E_+$) in the phase-space, then the trajectory emerged from that initial condition is travelled into the $E_+$ region when the excitation $F\sin(\omega t)$ takes negative values. Since the internal frequency of the oscillator and the excitation frequency are dissimilar, then the trajectories which are crossing the $E_+$ domain are trapped into the dissipative region for a long time, and occasionally comes out of it to make large excursions in the conservative region. This phenomenon is possible only when the initial conditions are originated or near the $E_{+}$ energy region. Nevertheless, the trajectories that are begun away from the dissipative region ($E_+$) have another basin of attraction with quasiperiodic dynamics. This is clearly illustrated in Fig.~\ref{fig7} in which the dissipative energy regions ($E_-, E_0$ and $E_+$) for different values of $F$ (as discussed in Fig.~\ref{fig6}) are plotted for comparison. An initial condition (marked as a filled triangle) with energy $E_1$ ($E_2>E_1>E_+$) originated near to the energy region $E_+$ is considered. The trajectory evolved from this initial condition is trapped inside the dissipative region ($E_{+}$) for a long time and occasionally come out of it, exhibiting EEs. Nevertheless, another initial condition (indicated as a filled circle) with energy $E_2$, started away from the energy region $E_+$ has quasiperiodic dynamics.

Next, the authors have considered another dynamical system, a microelectromechanical cantilever system model, to illustrate the parametric excitation induced EEs phenomenon.
%
\section{\label{sec3}Extreme events in microelectromechanical system}
The normalized dimensionless equation of motion of the MEMS model is given as\cite{ksuresh2018,liu2004, evans2014}
\begin{align}
\ddot{x}+\alpha \dot{x}+\gamma [1+F \cos(\omega t)]x-\frac{\beta^2}{(1-x)^{2}} = 0,
\label{eq2} 
\end{align}
where $x$ is the displacement of the movable cantilever structure concerning the $x$-axis. $\alpha$ is the strength of the damping force, $\gamma$ belongs to the stiffness constant of the structure, $\beta$ represents the strength of the nonlinear electrostatic actuation force, $F$ and $\omega$ are the amplitude and frequency of the parametric excitation, respectively. Equation (\ref{eq2}) was numerically integrated using the Runge-Kutta integration method with adaptive step size and to obtain the numerical solutions, the system parameters are fixed as follows: $\alpha$ = 0.71, $\gamma$ = 0.5, and $\beta$ = 0.32. The system (\ref{eq2}) has a discontinuous boundary due to the presence of singularity at $x=1$ and so it has two subspaces at $x>1$ and $x<1$\cite{ksuresh2018}. Systems with discontinuous boundaries are frequently encountered in mechanical \cite{liu2004, evans2014, geffert2017, li2016} and laser models \cite{chizhevsky1997, bonatto2017}. The nonlinear behavior of the system (\ref{eq2}) has already been studied in the literature (refer to the review article \cite{rhoads2010} and the references therein). In particular, theoretical analysis and experimental results on the dynamical behavior of a bistable MEMS oscillator have been investigated \cite{de2006}. Also, the influence of super-harmonic excitation in the MEMS model is studied, and the emergence of chaotic oscillations via period-doubling bifurcation is observed \cite{de2005}. In addition to that, the resonance property of the parametrically excited MEMS model without the nonlinear term ($\beta=0$ in Eq.~(\ref{eq2})) is studied which is used in microscale gyroscopes \cite{miller2008, gallacher2006}.

Further, in a recently reported study\cite{ksuresh2018}, the authors have incorporated the external force as an additive term in the autonomous MEMS model and demonstrated the development of EEs. However, the influence of parametric excitation to induce EEs in the MEMS model and its developing mechanism are still unknown. Therefore, in the present paper, the authors have considered the MEMS model with parametric excitation (forcing as a multiplicative term as given in Eq. (7)) and demonstrated the emergence of EEs as a function of the excitation parameters $F$ and $\omega$.

\begin{figure}[tbp]
	\centering
	\includegraphics[width=1.0\columnwidth]{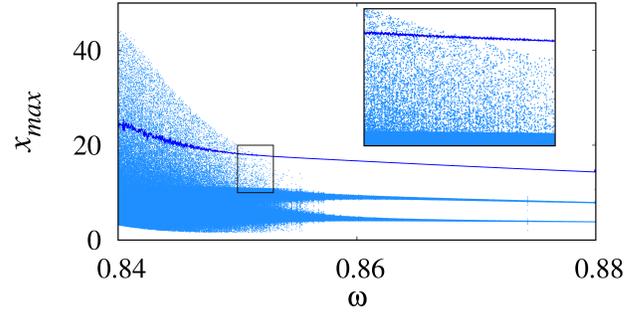}
	\caption{The one-parameter bifurcation diagram of the MEMS model (\ref{eq2}) as a function of parametric excitation amplitude $\omega\in(0.84,0.88)$. The other system parameters are fixed as $\alpha = 0.71$, $\gamma = 0.5$, $\beta = 0.32$, and $F = 1.48$. $H_T$ is plotted as a blue (dark gray) line. The inset depicts the magnified view of the rectangular part to show the critical value of $\omega$ in which the system exhibit EEs.} 
	\label{fig8}
\end{figure}

Without the parametric excitation ($F=0$ in Eq.~(\ref{eq2})), the system has three equilibrium points $X_{0} = (x_{0},0)$ and $X_{1,~2}=\left(\frac{2-x_{0}\pm\sqrt{(4-3x_{0})x_{0}}}{2}, 0\right)$ when $\gamma>27\beta^{2}/4$. Else, the system has only one fixed point of $X_{0}$. Here $x_{0}$ is the solution of the equation $x_{0}^{3}-2x_{0}^{2}+x_{0}-\beta^{2}/\gamma = 0$. For the above-chosen parameter values, the system has only one fixed point at $X_{0}$ = (1.38248, 0), which is a stable focus. If one incorporates the parametric excitation and slowly decreases its frequency from higher to lower values by fixing the amplitude of the excitation, then the system exhibits various dynamical states. These dynamical transitions can be realized by plotting the bifurcation diagram of the system (\ref{eq2}), which is depicted in Fig.~\ref{fig8}, shows the qualitative changes that occurred in the dynamical variable $x$ of the MEMS model as a function of $\omega\in[0.84, 0.88]$ for $F=1.48$. The authors define EEs in the system (\ref{eq2}) when the system variable $x$ exceeds four times ($n=4$) the standard deviation over mean peak value $\langle P\rangle$ in Eq. (\ref{eq1a}). The blue (dark gray) line in Fig.~\ref{fig8} indicates the threshold height $H_T$ calculated using Eq. (\ref{eq1a}) in which $\langle P\rangle$ is the mean peak value of the $x$ component of the MEMS model (\ref{eq2}), and $\sigma$ is the standard deviation of the $x$ variable. For $\omega\geq0.875$, the system shows periodic oscillations, and when the excitation reaches the value of $\omega=0.875$, the system displays small-amplitude chaos which is evident from Fig.~\ref{fig8}. The amplitude of the chaotic oscillations is then slowly increasing with decreasing the value of $\omega$. For $\omega=0.855$, there is a sudden emergence of large-amplitude chaotic bursts that are coexisted with the small-amplitude oscillations. However, these large peaks are below the threshold height $H_T$ and do not qualify as EEs. The amplitude of the large peaks is continuously increasing as of decreasing the frequency, and for $\omega=0.8513$, the large peaks are higher than $H_T$, corroborating the development of EEs in the system. The inset of Fig.~\ref{fig8} shows the magnified view of the small portion of the bifurcation diagram confirming the onset of EEs in the MEMS model.

\begin{figure}[tbp]
	\centering
	\includegraphics[width=1.0\columnwidth]{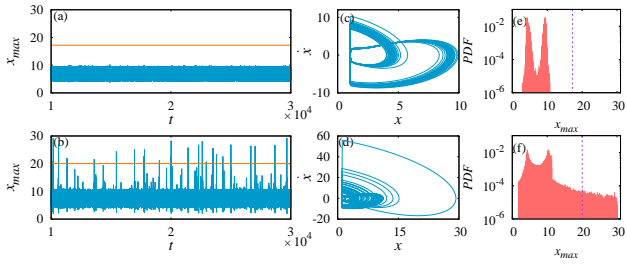}
	\caption{Time evolution of the system (\ref{eq2}) is plotted in the left vertical panel for different values of forcing frequency (a) $\omega = 0.856$ and (b) $\omega = 0.845$ for $F=1.48$. The corresponding attractor and the probability density function are depicted in middle [(c) and (d)] and right panels [(e) and (f)], respectively. The horizontal red (dark gray) line in the figures (a) and (b) and vertical dashed line in figures (e) and (f) represents the value of $H_T$ with $n=4$ in Eq.~(\ref{eq1a}).} 
	\label{fig9}
\end{figure}

The time evolution, phase portraits and the PDFs of the MEMS model are plotted in Fig.~\ref{fig9} for two different values of $\omega$. Figure~\ref{fig9}(a) depicts the time traces of the maxima of the system variable $x$ for $\omega=0.856$, shows the small-amplitude chaotic oscillations. The horizontal line represents the threshold height $H_T$. The corresponding phase-space plot is drawn in Fig.~\ref{fig9}(c) and the localized structure of the PDF plotted in Fig.~\ref{fig9}(e) also validating the small-amplitude oscillations of the system. The double-hump like the structure of the PDF in Fig.~\ref{fig9}(e) shows that the maxima of the system variable $x$ are confined in two different places of the phase-space, which is visible in the attractor (Fig.~\ref{fig9}(c)). If $\omega$ is decreased further, then the system exhibits intermittent large-amplitude oscillations along with the small-amplitude chaos as illustrated in Fig.~\ref{fig9}(b) for $\omega=0.845$ as a time series plot. From this figure, one can notice the emergence of frequent large peaks that are coexisted with the small-amplitude oscillations. Among them, only very few peaks are larger than the threshold height $H_T$, qualified as EEs. The large excursions of the orbit in Fig.~\ref{fig9}(d) corroborate the appearance of EEs. Moreover, the long-tail nature of the PDF in Fig.~\ref{fig9}(f) is also confirming the occurrence of EEs.

\begin{figure}[tbp]
	\centering
	\includegraphics[width=1.0\columnwidth]{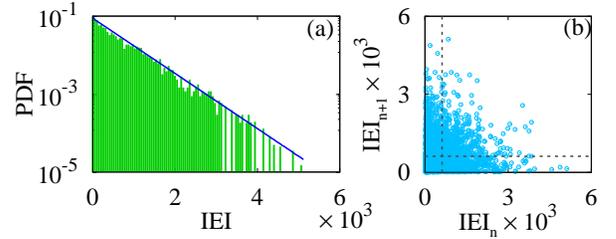}
	\caption{(a) IEI histogram and (b) JIEI plot of the MEMS model (\ref{eq2}) shows the Poisson-like distribution corroborating the emergence of EEs for $\omega=0.845$ and $F=1.48$.} 
	\label{fig10}
\end{figure}

The occasional occurrence of the EEs is again confirmed by estimating IEI and JIEI histograms of the MEMS model for $\omega=0.845$ (equivalent to Fig.~\ref{fig9}(b)) which are depicted in Figs.~\ref{fig10}(a) and \ref{fig10}(b), respectively. The PDF of the IEI is then fitted by $P(x)=\lambda e^{-\lambda x}$ with scaling parameter $\lambda=0.00163$ (marked as a blue (dark gray) line in Fig.~\ref{fig10}(a)) follows a Poisson-like distribution. Next, the negative exponential distribution of JIEI histogram is plotted in Fig.~\ref{fig10}(b) once again confirming the rare emergence of the large-amplitude oscillations. The mean value of the JIEI is shown as a dashed line in Fig.~\ref{fig10}(b).

\begin{figure}[tbp]
	\centering
	\includegraphics[width=1.0\columnwidth]{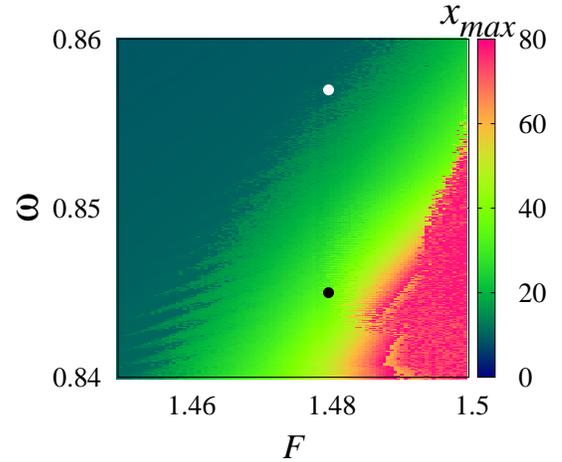}
	\caption{The two-parameter bifurcation diagram of the system (\ref{eq2}) as a function of the parametric excitation amplitude $F\in[1.35,1.5]$ and excitation frequency $\omega\in[0.84,0.86]$. The maximum peak value, $x_{max}$ is indicated in different colors. The dark green (gray) region shows the non-extreme event region where the peak values are below $H_T$. Other colored regions are indicating extreme event dynamics with different $x_{max}$ values. The white and black filled circles indicate the set of parameter values ($F$ and $\omega$) at which the time series of the Figs.~\ref{fig9}(a) and \ref{fig9}(b) are plotted.} 
	\label{fig11}
\end{figure}

A two-parameter phase diagram in the ($F,\omega$) plane is plotted in Fig.~\ref{fig11} to identify and distinguish the EEs region from the non-extreme event region. The dark green (gray) area in Fig.~\ref{fig11} represent the non-extreme event region where the maximum peak value of the system observable $x$ ($x_{max}$) is smaller than $H_T$. Other colored regions denote the large peaks, which are higher than $H_T$, corroborating the extreme event region. Different color gradients indicate different peak values of the MEMS model. One can also note that when the amplitude of the system is increased, then the corresponding standard deviation value ($\sigma$ in Eq.~(\ref{eq1a})) is also increases resulting in the increment of the threshold height $H_T$ accordingly. Therefore, as stated earlier, $H_T$ is calculated at each set of parameters. The white and black filled circles in Fig.~\ref{fig11} indicates the parameter values ($F$ and $\omega$) at which the time evolution plots in Figs.~\ref{fig9}(a) and Fig.~\ref{fig9}(b), respectively, are plotted.

\begin{figure}[tbp]
	\centering
	\includegraphics[width=1.0\columnwidth]{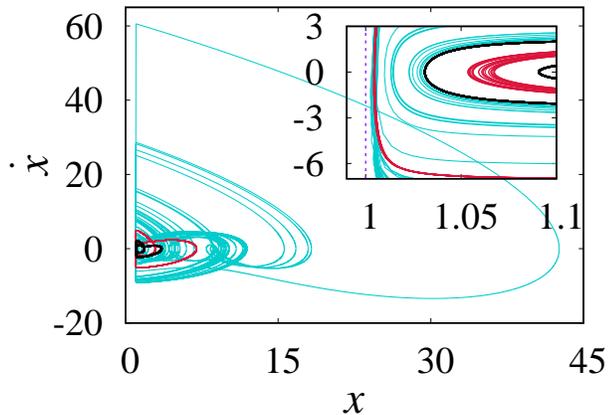}
	\caption{The phase-space diagram depicts the trajectories of the MEMS model (\ref{eq2}) for three different values of $\omega$. The black line represents the attractor for $\omega=0.875$. The bounded chaos is illustrated as a red (dark gray) line which is plotted for $\omega=0.856$. Finally, the bluish-green (light gray) trajectory represents the development of EEs for $\omega = 0.842$. The inset figure shows the maximized view of the trajectories near the discontinuous boundary $x=1$ and the vertical line denotes the discontinuous boundary.} 
	\label{fig12}
\end{figure}

It is important to emphasize here that, although the system is parametrically excited, the emerging mechanism of EEs is similar to the mechanism discussed in ref.\cite{ksuresh2018}. Nevertheless, the parameter space in which EEs occurred is entirely different. As discussed earlier, the MEMS model has a discontinuous boundary at $x=1$. This discontinuity induces the sliding motion of the trajectory in the phase-space. Hence, the trajectories which are entering into this sliding region undergo a stick-slip bifurcation. Therefore the trajectories are sliding over a long distance in the phase-space causes large-amplitude oscillations. The phase-space plot of the system (\ref{eq2}) is plotted in Fig.~\ref{fig12} for three different values of the excitation frequency ($\omega$) to elucidate this phenomenon. First, for $\omega=0.875$, the system exhibits periodic oscillations where the trajectories (black line in Fig.~\ref{fig12}) are far away from the discontinuous boundary (dashed line in the inset of Fig.~\ref{fig12}). Next, when one chooses $\omega=0.856$, the system show small-amplitude chaotic oscillations. However, still, the orbits are not close enough to the discontinuous boundary, and so the system does not undergo the stick-slip bifurcation. The red (dark gray) trajectories in the magnified image plotted as an inset of Fig.~\ref{fig12} confirming this case. Finally, for $\omega=0.842$, the trajectories are close enough to the discontinuous boundary results in a stick-slip bifurcation. Consequently, the trajectories undergo a large excursion in the phase-space leads to large-amplitude oscillations as depicted in Fig.~\ref{fig12} (bluish-green (light gray) lines). Since the system exhibit chaotic motion, the trajectories perhaps enter into the sliding region only at random intervals of time which is the reason for the rare and random occurrence of the large-amplitude oscillations.
%
\section{\label{sec4}Conclusion}
To consolidate the results, the authors have investigated the dynamics of the parametrically excited two nonlinear oscillators, namely Li\'enard type oscillator and MEMS model to demonstrate the emergence of EEs in terms of the parametric excitation parameters. In the Li\'enrd oscillator, the authors confirmed that the appearance of EEs occurred via two distinct dynamical routes: period-doubling and intermittency when varying the excitation frequency. The threshold height $H_T$ is calculated to distinguish EEs from the other dynamical states. Further, the long-tail behavior of the PDF additionally confirms the development of EEs in the Li\'enard oscillator. Furthermore, the authors also estimated the return time of the two successive EEs defined as IEI, for the long run to quantify the rare occurrence of EEs. They showed that the PDF of the IEI follows the Poisson-like distribution, which reconfirms the emergence of rare events. The possible mechanism responsible for the development of EEs in the Li\'enard oscillator is discussed based on the total energy of the system. Next, the authors have demonstrated the emergence of EEs in a parametrically excited MEMS model with discontinuous boundary and found that the occurrence of stick-slip bifurcation near the discontinuous boundary is the crucial mechanism for the appearance of EEs. The results presented here may be advantageous to understand the emergence of EEs occurred in macro, and micromechanical oscillators due to parametric excitation and also these bursting oscillations are found to be beneficial for energy harvesting applications \cite{cohen2012}.\\

\noindent {\bf AUTHOR'S CONTRIBUTIONS}\\
All authors contributed equally to this work.
\begin{acknowledgments}
The work of R. S. is supported by SERB-DST Fast Track scheme for young scientists under Grant No. YSS/2015/001645. The work of V. K. C. is sponsored by the SERB-DST-MATRICS Grant No. MTR/2018/000676 and CSIR EMR Grant No. 03(1444)/18/EMR-II.
\end{acknowledgments}
\vskip 10pt
\noindent {\bf DATA AVAILABILITY}\\
The data that support the findings of this study are available from the corresponding author upon reasonable request.\\

\end{document}